\documentstyle[12pt,epsf,epsfig,wrapfig]{article}
\begin{document}
\begin{center}
{\bfseries STAR RESULTS FROM POLARIZED PROTON COLLISIONS AT RHIC}
\vskip 5mm
L.C. Bland, {\it for the STAR collaboration}
\vskip 0.5mm
{\small
{\it
Brookhaven National Laboratory
}
}
\vskip 0.5mm
$\dag$ {\it
E-mail: bland@bnl.gov
}
\end{center}
\vskip 5mm
\begin{abstract}
This talk reports on progress from the first two years of polarized
proton collisions at the Relativistic Heavy Ion Collider (RHIC) at
Brookhaven National Laboratory (BNL).  STAR is one of the two large
experiments at RHIC.  It features large acceptance spanning a broad
range of rapidity.  The long-term goals of the STAR spin program are
to measure the gluon contribution to the proton's spin; determine the
contribution of specific quark flavors to the spin of the proton
through the study of spin observables for vector boson production, and
to determine the transversity structure function.  The study of
polarized proton collisions at $\sqrt{s}$=200 and 500 GeV is expected
to provide important insight into the spin structure of the proton,
revealing the contributions to the spin sum rule either from gluons or
from the orbital motion of the partons.  Selected STAR results from the
study of polarized proton collisions during the first two runs are
presented.
\end{abstract}
\vskip 8mm

\section{Introduction}

The parton model codifies our present understanding of the proton.
Viewed in the Breit frame, the proton consists of collinear quarks and
gluons each carrying a fraction of the proton's momentum, given by
Bjorken $x$.  Global analyses of many spin-independent high-energy
scattering observables ({\it e.g.}, Ref.~\cite{CTEQ}) provide details
of the Bjorken $x$ dependence of the quark and gluon densities.
Global analyses of polarized deep inelastic scattering experiments
demonstrate that the quarks alone cannot account for the proton's spin
(for a recent analysis, consult Ref.~\cite{BB} and references
therein).  To satisfy the proton's spin sum rule, either the gluons
must be polarized or there must be significant contributions to the
proton's spin from the orbital motion of its constituents.  If the
latter, then the transverse momentum of the quarks and gluons cannot
be ignored, as in the parton model.  One of the goals of the RHIC spin
program \cite{Bunce} is to determine whether gluons or parton orbital
motion account for the part of the proton's spin not carried by the
quarks.

The Relatistic Heavy Ion Collider (RHIC) at Brookhaven National
Laboratory is designed with the capability of accelerating and
colliding high-energy beams of polarized protons.  Measurements of
polarization observables for the inclusive production of hadrons,
hadronic jets, and photons in longitudinally polarized
$\vec{p}+\vec{p}$ collisions at $\sqrt{s}$=200 and 500 GeV will
address whether gluons contribute to the proton spin.  More exclusive
measurements, such as $\gamma$+jet final states, hold the promise of
determining the Bjorken $x$ dependence of gluon polarization within
the framework of leading-order pQCD \cite{EPIC}.  The extensions to
next-to-leading order that have already been made for inclusive
observables are still needed for these more exclusive final states.
Measurements of spin observables for $W^\pm$ and $Z^0$ production will
allow a decomposition of the flavor dependence of quark and antiquark
contributions to the proton's spin.  The study of transversely
polarized $p+p$ collisions at RHIC can provide information about quark
transversity ($\delta q(x) = q_\uparrow(x) - q_\downarrow(x)$) and
possibly some insights into orbital angular momentum of the partons.
The physics goals of the two large experiments (PHENIX and STAR) at
RHIC include the study of the proton's spin structure via measurements
of spin observables in polarized proton collisions.

The primary focus of the first three RHIC runs has been on the study of
Au+Au collisions and d+Au collisions to search for evidence of the
quark-gluon plasma.  Portions of the last two RHIC runs have been
committed to commissioning the accelerator for polarized proton
collisions, and to conduct the first studies of of $\vec{p}+\vec{p}$
collisions at $\sqrt{s}$=200 GeV.  Nine-week periods in RHIC run 2
(from December, 2001 to January, 2002) \cite{SPIN2002} and RHIC run 3
(from March, 2003 to May, 2003) were devoted to this purpose.  In run
2, only vertically polarized beams were available, since spin rotator
magnets had not yet been installed.  In run 3, spin rotator magnets
were in place at the STAR and PHENIX interaction regions and were
commissioned to provide the first collisions of longitudinally
polarized protons \cite{mackay}.  In run 2, an integrated luminosity
of 0.3 pb$^{-1}$ with an average beam polarization of 15\% was
delivered to STAR.  In run 3, 0.5 pb$^{-1}$ of vertically polarized
proton collisions and 0.4 pb$^{-1}$ of longitudinally polarized proton
collisions, with an average polarization of 25\%, was delivered to
STAR.

An important goal for run 2 was to identify processes that had
sizeable analyzing powers and yields.  Since the stable spin direction
of RHIC is vertical, special magnets either side of the STAR (and
PHENIX) interaction region (IR) are required to precess the spin to
make longitudinal polarization at the IR and then back again after the
IR.  Local polarimeters are required to establish that vertical and
radial polarization components of the colliding proton beams are small
when the rotator magnets are excited.  As described below, two such
processes were identified in measurements at STAR.  In run 3, a
primary objective was to embark on measurements of $A_{LL}$ for
mid-rapidity inclusive jet production, a process sensitive to gluon
polarization.  This talk reports on selected spin physics results from
the STAR experiment during the first two years of polarized proton
operations at RHIC.

\section{Cross sections and analyzing powers for large rapidity
pion production}

The E-704 experiment at Fermilab observed large analyzing powers
($A_N$) for pion production at large Feynman $x$ ($x_F > 0.3$) and
moderate transverse momentum (0.5 $< p_T <$2.0 GeV/c) using a tertiary
polarized proton beam in a fixed target experiment with total energy
in the center of mass equal to $\sqrt{s}$=20 GeV \cite{E704}.  Similar
behavior has been observed at even lower $\sqrt{s}$ values
\cite{low_ener}.  Large values of $A_N$ were unexpected because the
chiral properties of QCD result in only very small analyzing powers
for partonic scattering processes involving u and d quarks.
Subsequent theoretical work suggested that large $A_N$ for particle
production could arise via correlations between transverse momentum
($k_T$) and the spin in either distribution functions (the Sivers
effect \cite{Sivers,ABM}) or fragmentation functions (the Collins
effect \cite{Collins,Anselmino}).  It is also possible that explicit
higher twist effects can give rise to large $A_N$
\cite{EfremovTeryaev,QiuSterman,Koike}.  Models incorporating these
effects were developed and their parameters were adjusted to fit the
E704 results.  These models were subsequently extended to RHIC
collision energies, leading to theoretical expectations of large
analyzing powers for forward pion production in $\vec{p}+\vec{p}$
collisions at $\sqrt{s}$=200 GeV.

\begin{wrapfigure}{L}{8.2cm}
\mbox{\epsfig{figure=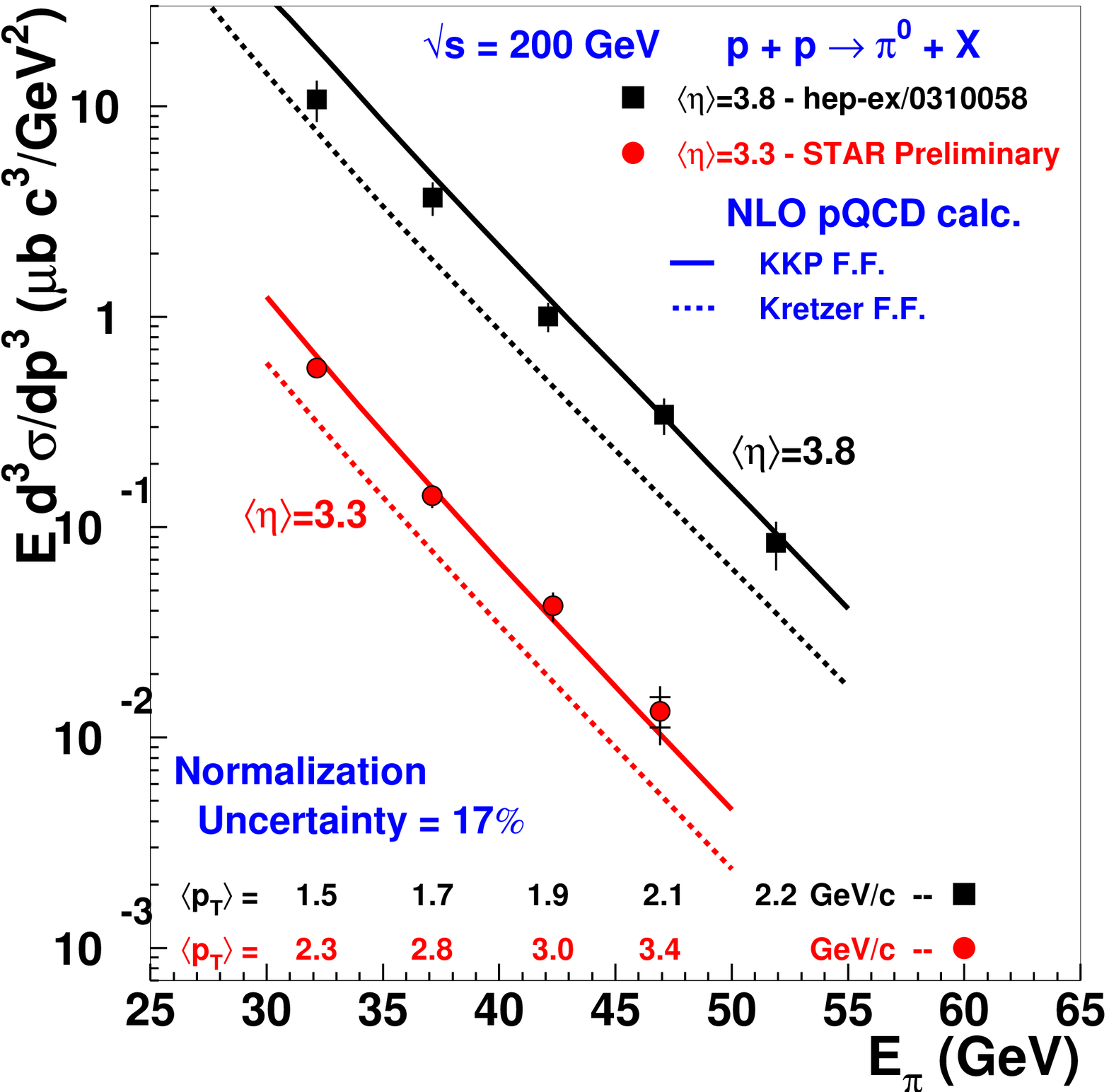,width=8.0cm,height=8.0cm}}
{\small{\bf Figure 1.} Invariant cross section for neutral pions
produced in p+p collisions at $\sqrt{s}$=200 GeV.  The cross section
points are shown versus the average energy of the pion, detected at
fixed pseudorapidity.}
\label{cross_section}
\end{wrapfigure}

The invariant cross section for a given process is an important
observable to help establish the dynamical origin of particle
production.  At low collision energies, it is generally believed that
large rapidity particle production arises from soft processes
collectively known as beam fragmentation.  Next-to-leading order
perturbative QCD calculations at $\sqrt{s}=20$ GeV underpredict
measured cross sections for pion production at large $x_F$ and small
$p_T$ by nearly an order of magnitude \cite{BS}.  Cross section
measurements at $\sqrt{s}$=200 GeV can help to clarify the relevant
dynamics for particles produced at large rapidity.  This is
particularly important to understand because of expectations that pion
production at large rapidity and moderate transverse momentum is in the
kinematic domain where gluon saturation effects in a heavy nucleus
will be manifest \cite{KKT}.

A prototype Forward $\pi^0$ Detector was used for measurements of
large $x_F$ $\pi^0$ production cross sections and analyzing powers in
the first polarized proton collision run at RHIC.  Details of the
measurements are reported elsewhere \cite{STAR_FPD}.  The results for
the invariant cross section measured are shown in
Fig.~1.  The data are compared with NLO pQCD
calculations evaluated with the CTEQ6M parton distribution functions \cite{CTEQ}
and equal renormalization and factorization scales set to $p_T$.  Two
sets of fragmentation functions are used.  The ratio of the computed
cross sections from the two fragmentation functions are similar to
what is observed at mid-rapidity \cite{PHENIX}.  The agreement between
the NLO pQCD calculation and the data is comparable to what is
observed at mid-rapidity.  Even though the transverse momenta are
small, the agreement suggests that particle production at large
rapidity in p+p collisions at $\sqrt{s}$=200 GeV is dominated by
partonic scattering, rather than soft mechanisms presumed responsible
for beam fragmentation.

\begin{figure}
\centerline{\epsfig{figure=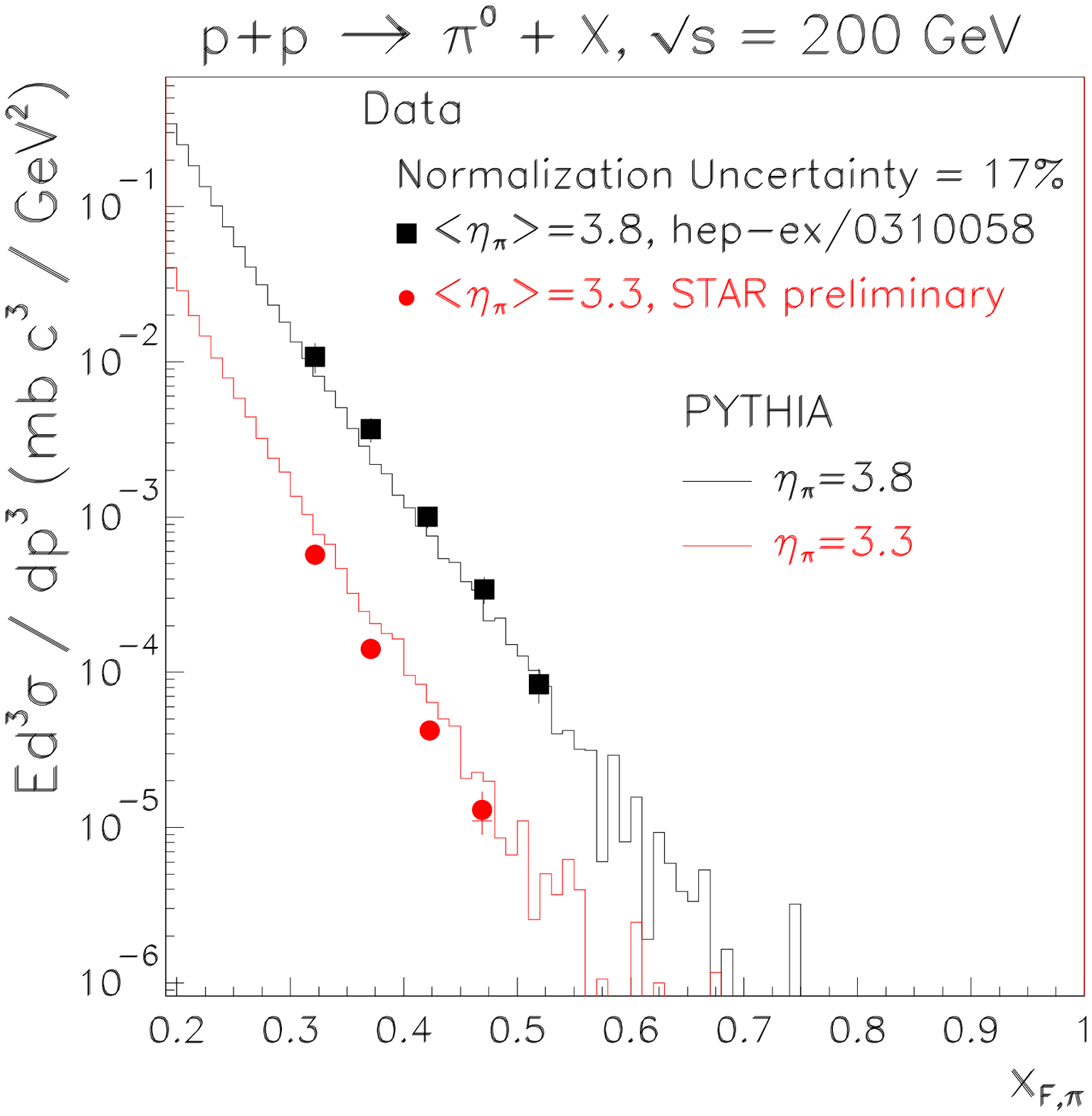,width=7.8cm,height=7.8cm}~\epsfig{figure=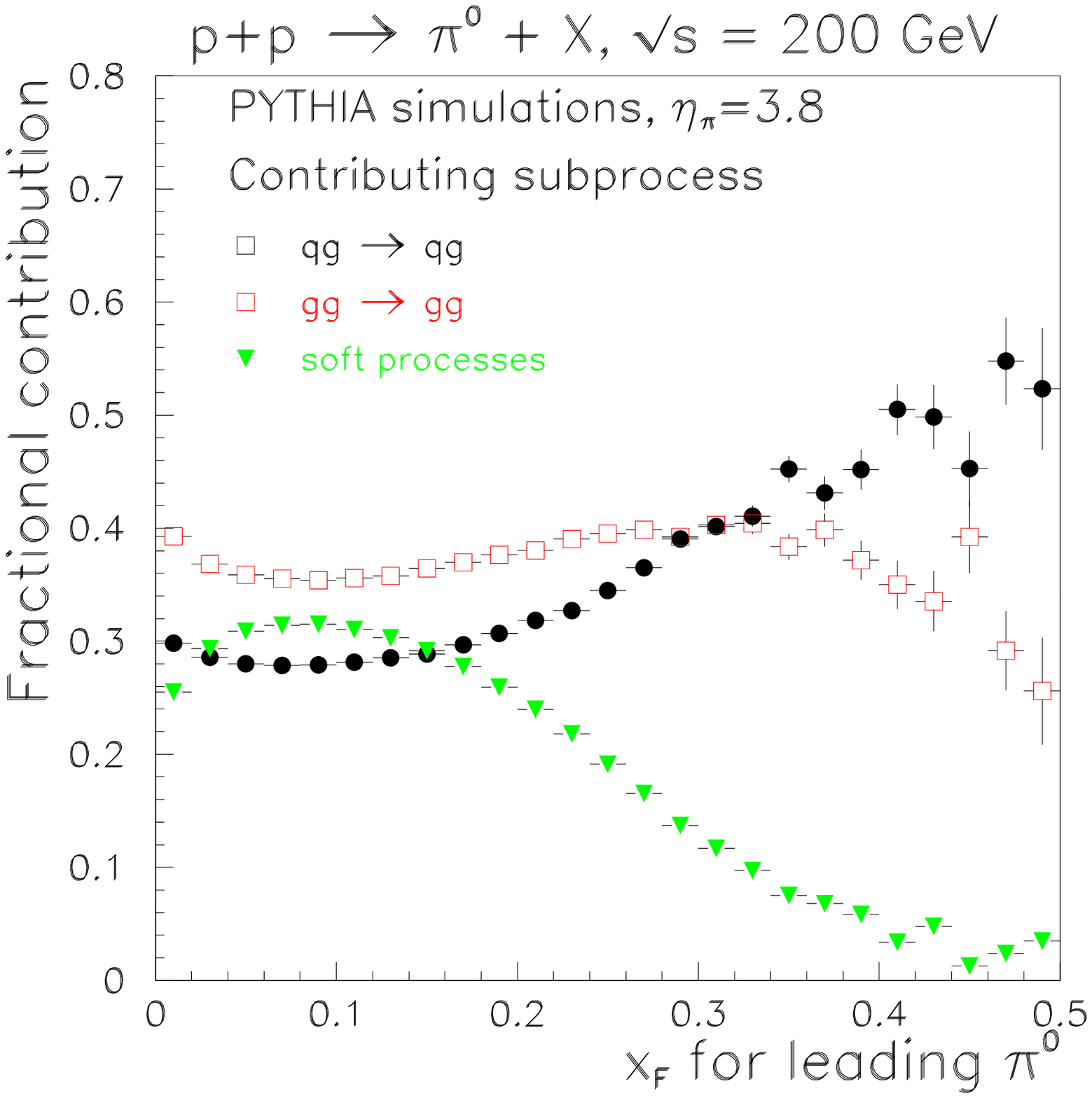,width=7.8cm,height=7.8cm}}
\medskip
{\small{\bf Figure 2.} (Left) PYTHIA \cite{PYTHIA} prediction for
large rapidity neutral pion production compared to measurements.
(Right) Fractional contribution of different PYTHIA
subprocesses to the production of neutral pions at $\eta$=3.8 from
$p+p$ collisions at $\sqrt{s}=$200 GeV.}
\label{PYTHIA_process}
\end{figure}

In addition to agreeing with NLO pQCD, the measured neutral pion
production cross sections at pseudorapidities $<\eta>=3.3$ and 3.8
also agree with results from the PYTHIA Monte Carlo generator
\cite{PYTHIA} (left panel of Fig.~2).  This is a non-trivial result,
since small angle particle production measurements are not available
at $\sqrt{s}$=200 GeV, meaning that the PYTHIA simulation is a
prediction, rather than resulting from tuning of its variables.  Given
that agreement, it is possible to establish the relative contributions
from the multiple different subprocesses.  For $x_F>0.3$, most of the
contributions come from initial states involving a large-$x$ quark and
a low-$x$ gluon.  Initial-state parton showers, where the large-$x$
quark splits into a quark+gluon, and the gluon subsequently undergoes
a hard interaction with a low-$x$ gluon from the other proton, is
responsible for a significant fraction of the forward pion yield
(right panel of Fig.~2).

Concurrent with the cross section measurements, the analyzing power
for neutral pion production at $<\eta>=3.8$ was measured for
$\vec{p}+\vec{p}$ collisions at $\sqrt{s}$=200 GeV in RHIC run 2
(Fig.~3).  At $<\eta>=3.3$, the cross section is an order of magnitude
smaller for the same $x_F$, hence the statistics for the $A_N$
measurement were insufficient.  As for the E704 data, the analyzing
power is found to be an increasing function of energy, proportional to
Feynman $x$.  Unlike the situation for E704, the concurrently measured
cross section is found to be in agreement with NLO pQCD.  In Fig.~3,
the $A_N$ measurements are compared to model calculations
incorporating spin-correlated $k_T$ effects in the distribution
function (Sivers effect) \cite{ABM}, a non-zero transversity
distribution and spin-correlated $k_T$ effects in the fragmentation
function (Collins effect) \cite{Anselmino}, and higher-twist effects
\cite{QiuSterman,Koike}.  All of these models fit the E-704 data
\cite{E704} and were predictions made by extrapolation from the lower
$\sqrt{s}$ to RHIC energies, evaluated at a fixed $p_T$ of 1.5 GeV/c.

\begin{wrapfigure}{L}{8.2cm}
\mbox{\epsfig{figure=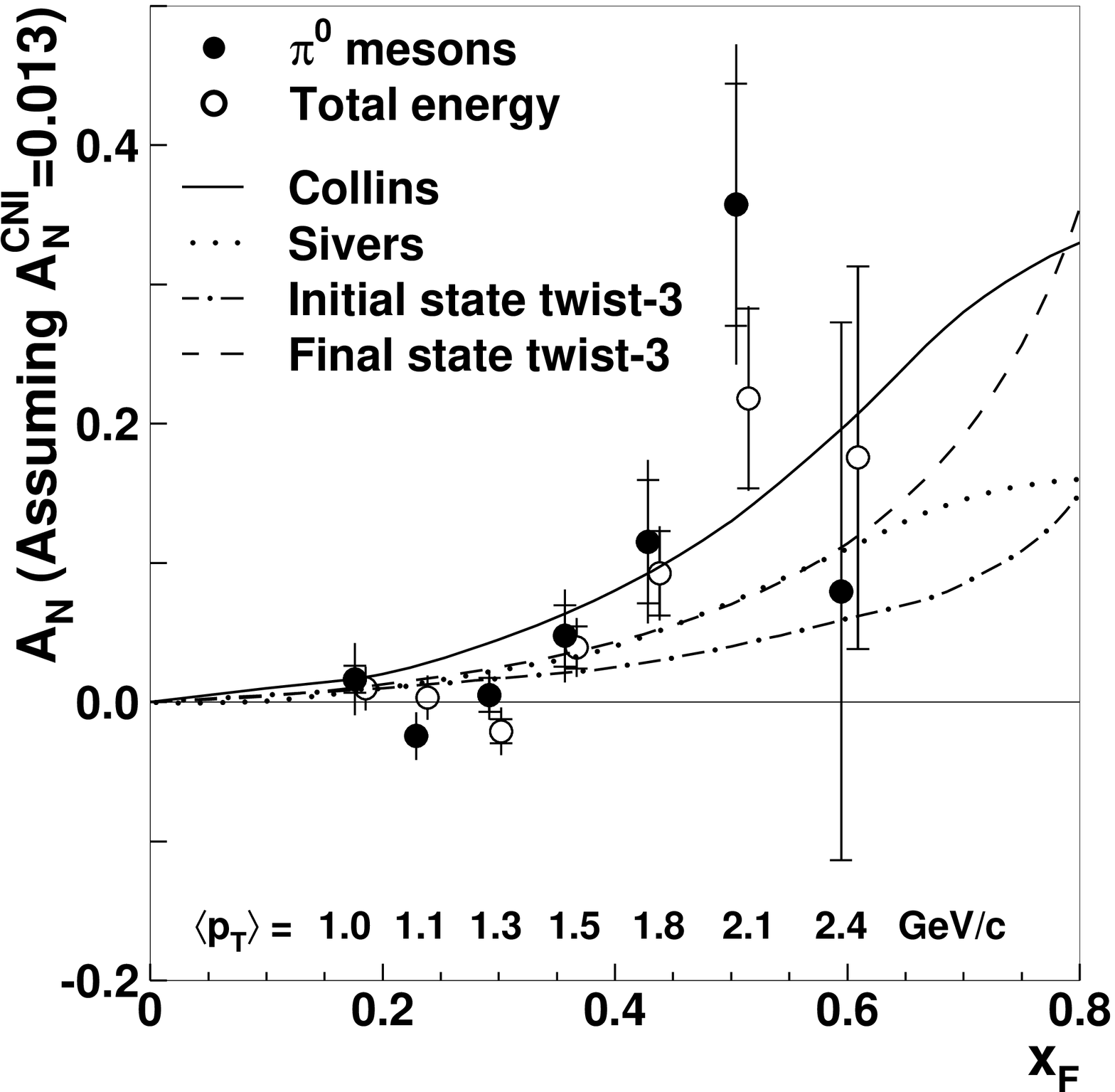,width=7.8cm,height=7.8cm}}
{\small{\bf Figure 3.} Analyzing power for neutral pions
produced in p+p collisions at $\sqrt{s}$=200 GeV.  The theoretical
curves are described in the text.}
\medskip
\label{analyzing_power}
\end{wrapfigure}

The magnitude of the analyzing power has two uncertainties, both
related to knowledge of the beam polarization.  At the RHIC injection
energy of 24.3 GeV knowledge of the beam polarization is presently limited by
calibration experiments to $\pm30$\% \cite{CNI}.  Relative
determinations of the polarization of the RHIC beams rely on the
measurement of the azimuthal distribution of recoil carbon ions
emerging near 90$^\circ$ scattering angles from a ultra-thin ribbon
target inserted into the RHIC beam.  The kinematics correspond to
proton-carbon elastic scattering at small momentum transfer ($0.006 <
|t| < 0.030$ (GeV/c)$^2$).  The analyzing power for this reaction
arises from the interference of the calculable Coulomb amplitude and
the unknown strong interaction amplitude; hence, polarimetry in RHIC
is based on Coulomb Nuclear Interference (CNI).  Dependence on the
proton beam momentum is, at present, inferred from calculation
\cite{Trueman_Kopeliovich} to be small.  This is consistent with the
observation that the measured asymmetry
($\epsilon_{CNI}=P_{beam}\times A_N^{CNI}$) was nearly the same at injection
energy and at 100 GeV for many fills.  Since the beam acceleration
process is unlikely to increase the beam polarization ($P_{beam}$),
this suggests that $A_N^{CNI}$ at 100 GeV is no smaller than at the
injection energy.  So, the data in Fig.~3 assumes that the
$A_N$ for the CNI polarimeter at proton beam energy of 100 GeV is the
same as measured values at lower energy \cite{Tojo}.  Measurements of
$\vec{p}+\vec{p}$ elastic scattering using a polarized gas jet target
are expected to provide a 10\% determination of the beam polarization
in RHIC run 4 \cite{Bravar}.

The precision of the measurements of forward $\pi^0$ production in
$\vec{p}+\vec{p}$ collisions at $\sqrt{s}$=200 GeV is, at present,
insufficient to discriminate the various models.  Nonetheless, it can
be concluded that large analyzing powers persist to high collision
energies as anticipated by these models.  Furthermore, the particle
production appears to be dominated by partonic scattering, as
evidenced by the agreement of the cross section with NLO pQCD
calculations.  It is interesting to note that all of the models were
adjusted to fit the E-704 data, but have different $x_F$ dependences
at $\sqrt{s}=200$ GeV.  Hence, higher precision measurements may lead
to an improved understanding of the dynamics responsible for these
spin effects.  More direct measurements, such as correlation of the
spin effects with the Collins angle or correlation of the spin effects
with a fully reconstructed forward jet, are expected to provide a more
definitive method of establishing the dynamical origins of the large
analyzing power.  To that end, STAR developed and partially installed
a Forward $\pi^0$ Detector prior to the start of RHIC run 3, as shown
in Fig.~4.  Significant data samples were obtained triggering STAR by
large energy deposition in the FPD.  Those samples will allow study of
particle correlations in $\vec{p}+\vec{p}$ and $d+Au$ reactions at
$\sqrt{s_{NN}}$=200 GeV.  Furthermore, at the RHIC design luminosity,
the analyzing power for forward neutral pion production can be
employed as a local polarimeter for tuning and monitoring the spin
rotator magnets.

\begin{figure}
\centerline{\epsfig{figure=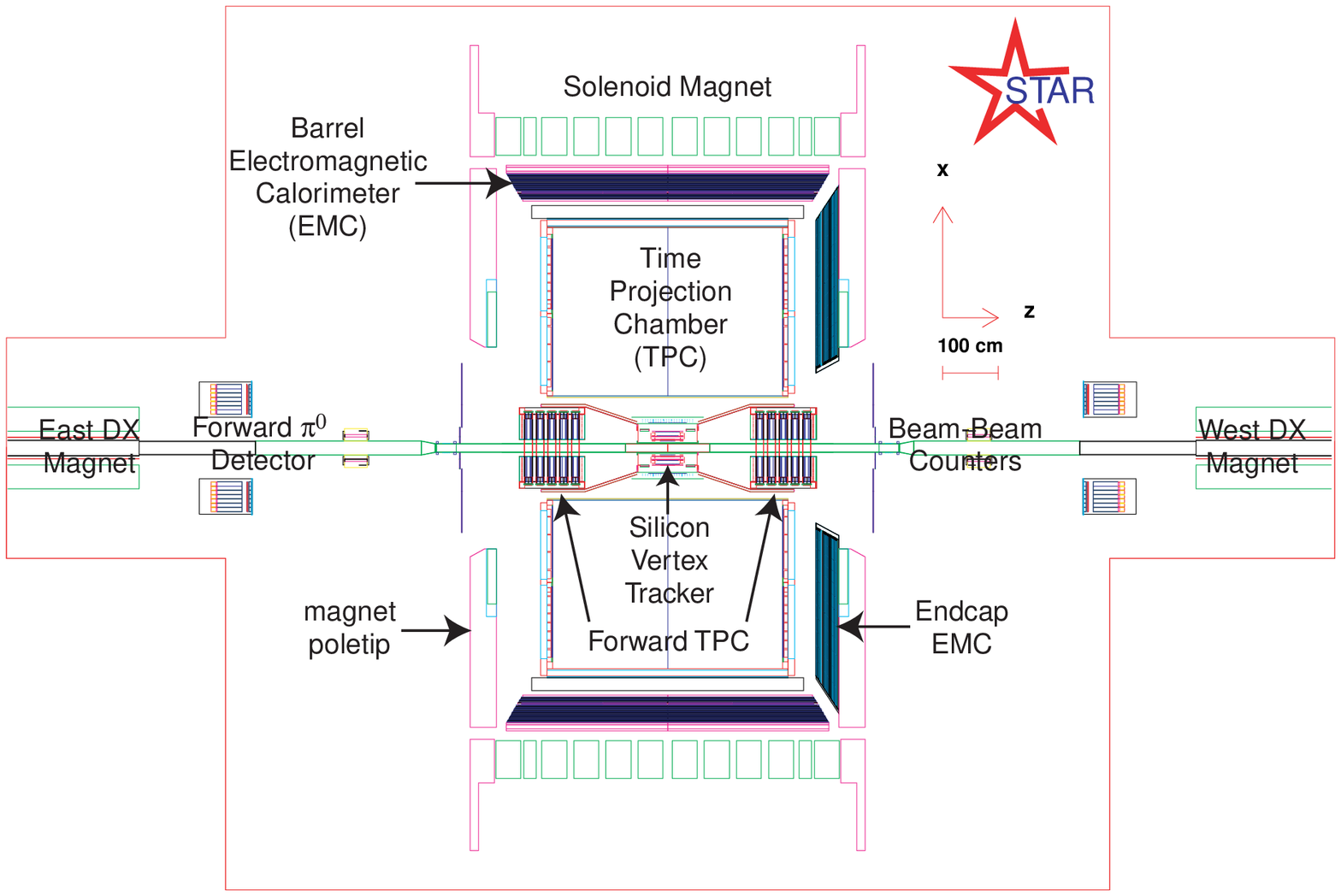,height=3.10in}}
\medskip
{\small{\bf Figure 4.} Layout of the STAR Wide Angle Hall showing the
completed barrel EMC, endcap EMC and FPD.  For run-3, the west half of
the barrel EMC and the lower half of the endcap EMC were in place.
All FPD components except the west-north calorimeter were in place for
run-3.  Note the difference between transverse ($x$) and longitudinal
($z$) dimensions.}
\label{run3}
\end{figure}

\section{Analyzing Powers for Large Rapidity Charged Particle Production}

The STAR beam-beam counters (BBC) are scintillator annuli that span
the pseudorapidity interval $2.2<|\eta|<5.0$.  The annuli are mounted
on each poletip of the STAR magnet and are separated by a distance
along the beam equal to 7.4 m.  Each annulus is tiled by optically
isolated hexagonal elements, as shown in Fig.~5.  Large tiles
cover the smaller $\eta$ range and small tiles the larger $\eta$ range.
Two rings of regular hexagonal cells, whose size is defined by an
inscribed circle of diameter 9.6 cm, provide azimuthally symmetric
coverage for the interval $3.4<|\eta|<5.0$.  These small tiles serve
as a minimum bias trigger for $p+p$ collisions, a luminosity monitor
and a local polarimeter, as described below.

\begin{wrapfigure}{L}{9.0cm}
\mbox{\epsfig{figure=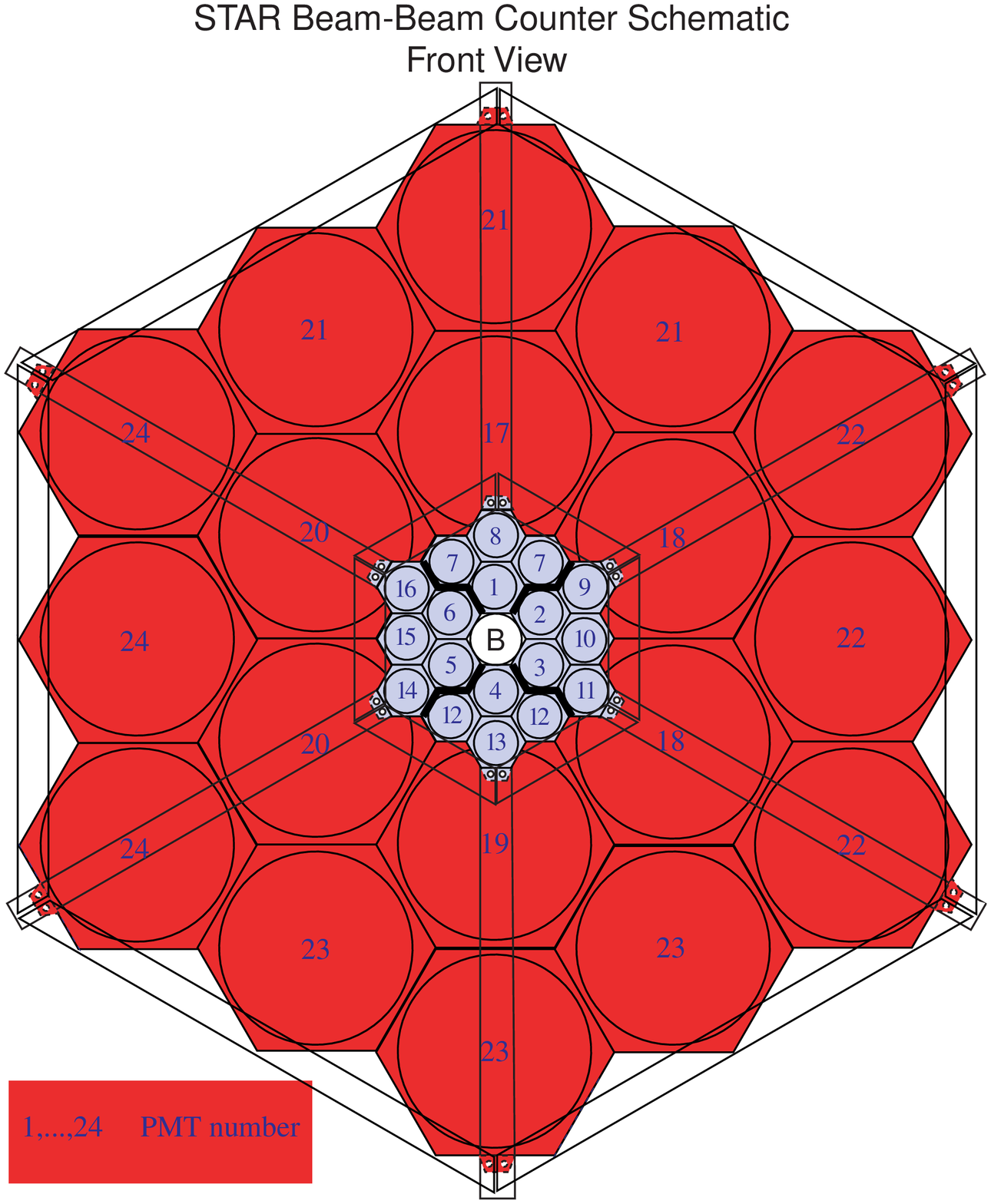,height=9.5cm}}\\
{\small{\bf Figure 5.} The STAR beam-beam counter
as seen looking towards the interaction point
from outside of the STAR magnet.  The large tiles are exactly four
times the size as the small tiles.  For spin-dependent analyses, the
small-tile annulus is divided into nearly equal quadrants, as
indicated.} 
\label{bbc}
\end{wrapfigure}

Each BBC phototube provides a summed charge proportional to energy
deposition and, for the small tiles, a time of arrival signal.  The
energy deposition is dominated by the energy loss of high energy
charged particles through the 1-cm thickness of plastic scintillator.
Collisions between the beams are discriminated from single beam backgrounds using
the timing signals.  A minimum bias trigger for $p+p$ collisions is
derived by requiring at least one hit on both sides of STAR with valid
timing.  The rate of these events has been calibrated as a luminosity
monitor.  Measurements of the trigger rate as the two beams are
scanned relative to each other determine the beam size.  Combining
that information with an independent measurement of the number of ions
in each ring determines the luminosity \cite{vanderMeer}.  Knowledge
of the luminosity then allows a determination of the total cross
section observed by the BBC.  This cross section is measured to be
$26.1\pm0.2$(stat.)$\pm1.8$(syst.) mb \cite{STARpp}, corresponding to
$87\pm8$\% of the inelastic, non-singly diffractive cross section.
This measurement is consistent with simulations based on PYTHIA
\cite{PYTHIA} and GEANT \cite{Kiryluk}.  The simulations provide a
good description of the overall detector response, the multiplicity of
hits from p+p collisions and the azimuthal and radial distributions of
the hits recorded in the minimum bias configuration.  The energy
deposited on one side of STAR corresponds to, on average, five charged
particles passing through the BBC small tile annulus for each p+p collision.

By correlating the azimuthal topology of hits in the BBC with
vertical beam polarization, a significant analyzing power is observed
for forward particle production at positive $\eta$ (relative to
the polarized proton beam).  Backward particle production, relative to
the polarized proton beam, is observed to have no analyzing power.
The azimuthal angle to associate with a given event is based on the
topology of hits.  As indicated in Fig.~5, the BBC annulus is divided into
quadrants for the analysis.  To analyze the vertical component of
polarization, the following spin asymmetry ($\epsilon$) is formed:
\begin{equation}
\epsilon=P_{beam}\times A_N =
{\sqrt{(L\cdot{\overline R})_\uparrow \times (R\cdot{\overline L})_\downarrow} -
\sqrt{(L\cdot{\overline R})_\downarrow \times (R\cdot{\overline L})_\uparrow}
\over
\sqrt{(L\cdot{\overline R})_\uparrow \times (R\cdot{\overline L})_\downarrow} + 
\sqrt{(L\cdot{\overline R})_\downarrow \times (R\cdot{\overline L})_\uparrow}}.
\label{spin_asym}
\end{equation}
Based on Fig.~5, Boolean expressions for the symbols are
$L$=(5+6+14+15+16) and $R$=(2+3+9+10+11), where the numbers refer to a
given phototube producing a sufficient pulse to exceed the
discriminator threshold for the event.  The symbols ${\overline
R}({\overline L})$ refer to the condition of no hits in the
corresponding phototubes, imposed to avoid ambiguities in the
azimuthal angle to assign to the event.  In Eqn.~\ref{spin_asym},
$\uparrow(\downarrow)$ refer to the direction of the polarization of
the particular bunch crossing.  As described elsewhere
\cite{Bunce,SPIN2002}, polarized proton operation in RHIC involves an
injection pattern of bunches with different polarization directions.
A similar analysis is applied to the top/bottom quadrants of the BBC.
For purely vertical polarization, these spin dependent asymmetries
must be zero.  Spin asymmetries from the top/bottom quadrants measure
radial polarization components of the beams at the interaction region.

A scaler system \cite{Crawford}, that counts at the RHIC
bunch-crossing frequency (9.35 MHz), facilitates high-statistics
measurements of spin asymmetries with the BBC.  During run-3, more than
1$\times 10^{10}$ BBC minimum bias events were recorded by the scaler
system with transverse polarization, and a comparable number of events
were recorded with longitudinal polarization.  Each scaler board has
24 input bits that serve as a memory address to a counter.  Seven of
the bits are reserved for the identification of the bunch crossing.
RHIC can have up to 120 bunch crossings separated by 107 ns.  On the
scaler board, one bit is reserved for the minimum bias trigger, and
the other 16 are driven by the discriminator outputs that establish if
individual phototubes of the BBC are above threshold (Fig.~5).
The resulting data are analyzed using Eqn.~\ref{spin_asym}, with the
spin direction determined from the polarization pattern used to inject
beam bunches into RHIC, as identified by the bunch crossing number.

The magnitude of the analyzing power observed with the BBC for
$\vec{p}+\vec{p}$ collisions at $\sqrt{s}=200$ GeV is approximately
half that observed by the CNI polarimeter ($A_N^{CNI}=0.013\pm 0.004$
\cite{CNI}), as determined by correlations of the spin dependent
asymmetry from the BBC with results from the CNI polarimeter from
measurements conducted at regular intervals through a RHIC store
\cite{Kiryluk}.  The analyzing power is found to have a strong
pseudorapidity dependence.  When hits are required on only the inner
ring of small tiles ($3.9 < \eta < 5.0$), the analyzing power is found
to be nonzero.  When hits are required on the outer ring of small
tiles ($3.4 < \eta < 3.9$), and no hits are present on the inner ring,
the analyzing power is found to be zero.

\begin{figure}
\centerline{\epsfig{figure=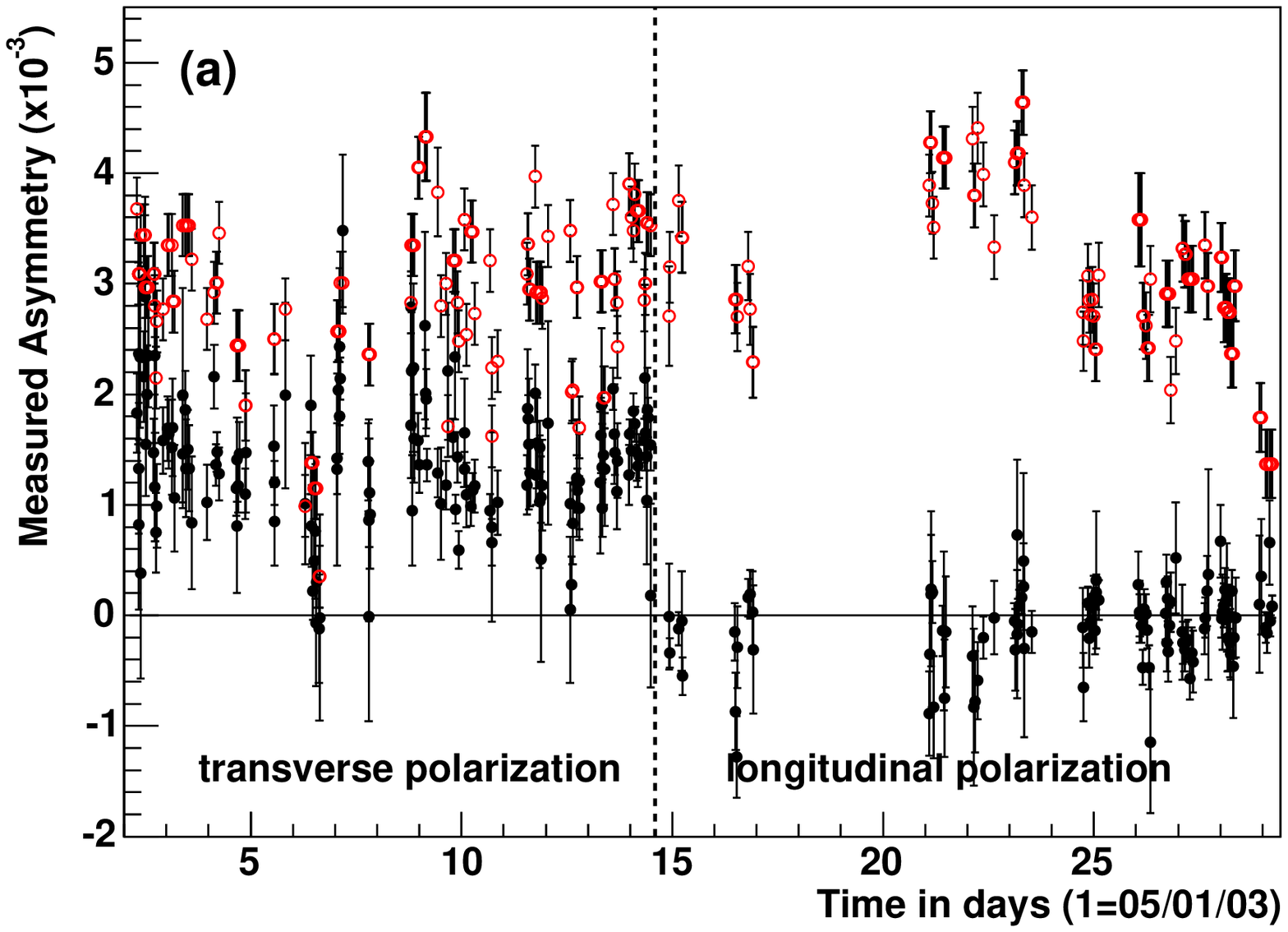,height=2.4in}
\epsfig{figure=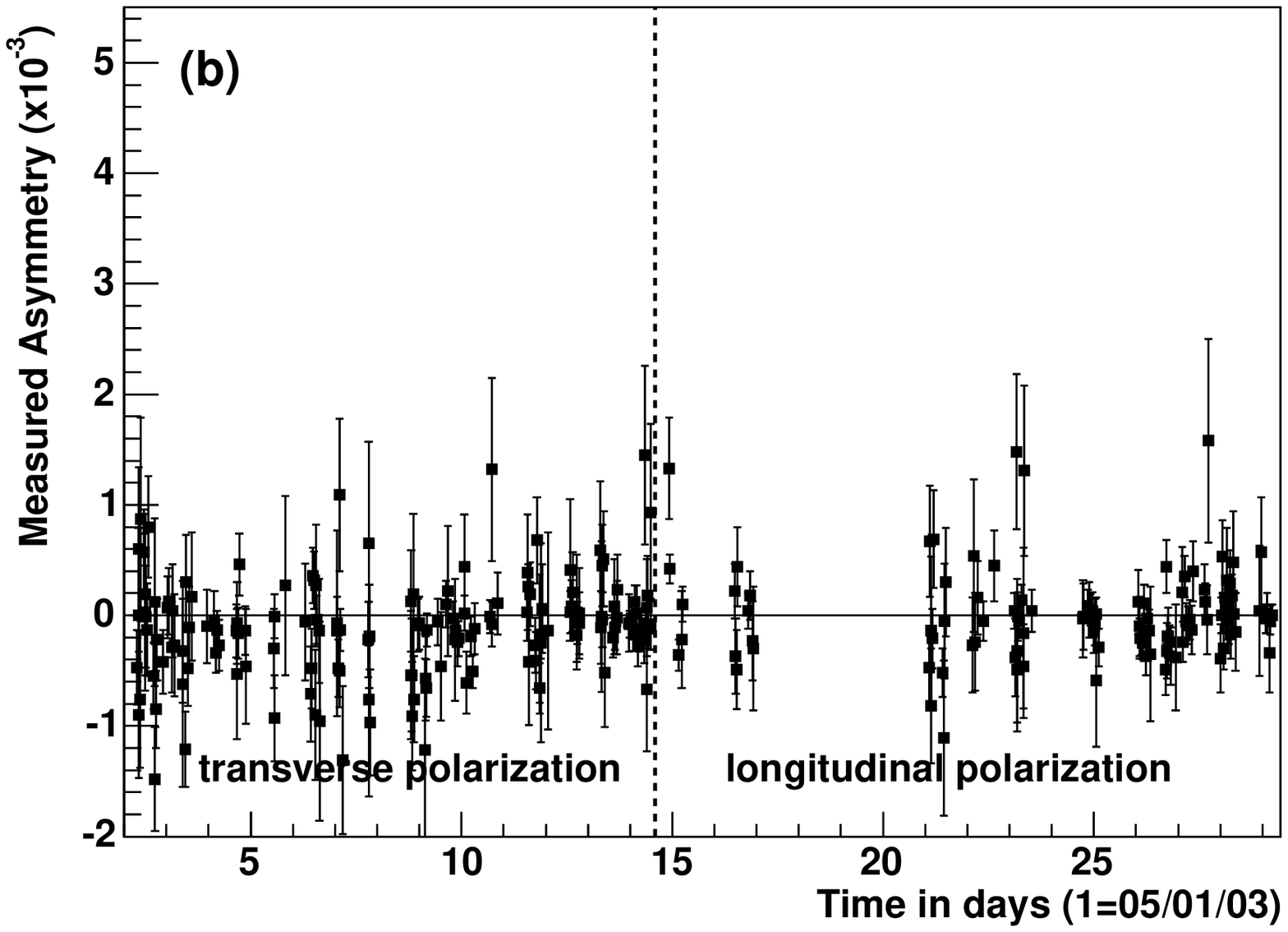,height=2.4in}}
\medskip
{\small{\bf Figure 6.} (a) The closed symbols are the raw single spin
asymmetries ($P_{beam}\times A_N$) for charged hadrons produced in the
pseudorapidity interval $3.4<\eta<5.0$ from polarized proton
collisions at $\sqrt{s}=200$ GeV, shown as a function of time.  The
time dependence results from variations in the beam polarization.  
The open symbols are the raw single spin asymmetries measured by the
RHIC CNI polarimeter.  The dashed line indicates when the STAR spin
rotator magnets were turned on.  This results in longitudinal
polarization at the STAR interaction region, thereby making the spin
asymmetry at STAR zero, while the CNI polarimeter results demonstrate
the beam remains polarized. (b) Results from top/bottom spin
asymmetries from the BBC as a function of time.  This demonstrates
that the spin rotator magnets stably produce longitudinal polarization
at the STAR interaction region.}
\label{bbc_asym}
\end{figure}

The dynamical origin of the spin effects observed with the BBC is
presently unknown.  They may be related to the E704 results \cite{E704},
since, based on simulations, $\pi^\pm$ mesons produced at large
Feynman $x$ are expected to be part of the charged particle flux
responsible for the hits observed in the BBC, but $x_F$ cannot be
measured by this detector.  Furthermore, at the
largest pseudorapidities, more $\pi^+$ are expected to be produced
than $\pi^-$ because of the proton's valence quark structure.  But
proving this from the BBC data alone is not possible, because neither
the particle's energy, its charge sign, nor its identity are
determined in the measurements.  Measurement of the $x_F$ dependence
of $A_N$ for charged pions would be of interest and is naturally suited to the
Brahms spectrometer \cite{Brahms}, likely restricted to scattering angles
that are larger than those from the BBC.

Even though the dynamics responsible for the spin effects observed
with the BBC are not known, the non-zero analyzing power coupled with
the rate capabilities of the scaler boards provides for a local
polarimeter that has very good statistical figure of merit.  At a
luminosity of $5\times 10^{30} {\rm cm}^{-2}{\rm sec}^{-1}$ and beam
polarization equal to 25\%, the direction of the polarization vector
for each beam can be established to 10$^\circ$ with a statistical
accuracy of 3 standard deviations within 30 minutes.
This capability proved useful in confirming that the STAR
spin rotator magnets were properly tuned to provide longitudinal
polarization at the STAR interaction region.  Furthermore, this local
polarimeter provides a continuous monitor of transverse polarization
components, as shown in Fig.~6 during operation with
longitudinally polarized beams. 

\section{Summary of present status and future plans}

The first collisions of spin polarized proton beams at $\sqrt{s}$=200
GeV have occurred at RHIC.  The STAR experiment has established that
the large analyzing powers observed for neutral pion production at
$\sqrt{s}=20$ GeV persist to the higher collision energy.  The cross
section for large rapidity neutral pion production is found to be in
agreement with NLO pQCD calculations, and also leading-order pQCD,
supplemented by the parton shower model \cite{PYTHIA}.  In addition, a
non-zero analyzing power was observed for inclusive charged particle
production in the pseudorapidity interval $3.4<\eta<5.0$.  This
analyzing power has provided STAR with a high figure-of-merit local
polarimeter, sensitive to transverse polarization components in the
colliding proton beams.

In RHIC run 3, STAR acquired the first data for mid-rapidity jet
production with both colliding beams longitudinally polarized.
Analysis of this data is underway.  In the near-term future, the plan
is to acquire a significantly larger inclusive jet data sample, with
improved beam polarization.  In the longer term, larger luminosity and
improved polarization will permit the start of spin observables for
direct photon measurements at $\sqrt{s}$=200 GeV.  Further runs with
transverse polarization are also under consideration.



\begin{thebibliography}{99}
\bibitem{CTEQ}
J.~Pumplin {\it et al.}, J. High Energy Phys. {\bf 0207}, 012 (2002).
\bibitem{BB}
J.~Bl\"umlein and H.~B\"ottcher, Nucl. Phys. {\bf B636},225 (2002).
\bibitem{Bunce}
Gerry Bunce, Naohito Saito, Jacques Soffer and Werner Vogelsang, Ann.\
Rev.\ Nucl.\ Part.\ Sci.\ {\bf 50}, 525 (2000).
\bibitem{EPIC}
L.~C.~Bland, in {\it Physics with a High Luminosity Polarized Electron
Ion Collider}, eds. L.~C.~Bland, T.~Londergan and A.~Szczepaniak
(World Scientific, Singapore, 2000).  Also available at
hep-ex/9907058.
\bibitem{SPIN2002}
L.~C.~Bland, AIP Conf. Proc. {\bf 675}, 98 (2003) and hep-ex/0212013. 
\bibitem{mackay}
W.~W.~MacKay {\it et al.}, ``Comissioning Spin Rotators in RHIC'', in
{\it Proceedings of the 2003 Particle Accelerator Conference}, p. 1697 (2003).
\bibitem{E704}
D.~L.~Adams {\it et al.}, Phys. Lett. B {\bf 261}, 201 (1991); {\bf
264}, 462 (1991).
\bibitem{low_ener}
W.~H.~Dragoset {\it et al.}, Phys. Rev. D {\bf 18}, 3939 (1978);
S. Saroff {\it et al.}, Phys. Rev. Lett. {\bf 64}, 995 (1990);
B.~E.~Bonner {\it et al.} Phys. Rev. D {\bf 41}, 13 (1990); K.~Krueger
{\it et al.}, Phys. Lett. B {\bf 459}, 412 (1999);
C.~E.~Allgower, {\it et al.}, Phys. Rev. D {\bf 65}, 092008 (2002).
\bibitem{Sivers}
D.~Sivers, Phys. Rev. D {\bf 41}, 83 (1990); {\bf 43} 261 (1991).
\bibitem{ABM}
M.~Anselmino, M.~Boglione and F. Murgia, Phys. Lett. B {\bf 362}, 164
(1995); M.~Anselmino and F.~Murgia, {\it ibid.} {\bf 442}, 470 (1998);
U.~D'Alesio and F.~Murgia, AIP Conf. Proc. {\bf 675}, 469 (2003).
\bibitem{Collins}
J.~Collins, Nucl.~Phys.~ {\bf B396}, 161 (1993).
\bibitem{Anselmino}
M.~Anselmino, M.~Boglione, and F.~Murgia, Phys. Rev. D {\bf 60},
054027 (1999); M.~Boglione and E.~Leader, Phys. Rev. D {\bf 61},
114001 (2000).
\bibitem{EfremovTeryaev}
A.~V.~Efremov and O.~V.~Teryaev, Phys. Lett. {\bf 150B}, 383 (1985).
\bibitem{QiuSterman}
J.~Qiu and G.~Sterman, Phys. Rev. D {\bf 59}, 014004 (1998).
\bibitem{Koike}
Y.~Koike, AIP Conf. Proc. {\bf 675}, 449 (2003).
\bibitem{BS}
Claude Bourrely and Jacques Soffer, hep-ph/0311110 (2003).
\bibitem{KKT}
Dmitri Kharzeev, Yuri V. Kovchegov and Kirill Tuchin, Phys. Rev. D
{\bf 68}, 094013 (2003).
\bibitem{STAR_FPD}
J.~Adams, et al (STAR collaboration), submitted to Phys. Rev. Lett.,
hep-ex/0310058. 
\bibitem{PHENIX}
S.~S.~Adler {\it et al.}, Phys. Rev. Lett. {\bf 91}, 241803 (2003).
\bibitem{PYTHIA}
T.~Sj\"{o}strand, Comp. Phys. Commun. {\bf 82}, 74 (1994).
\bibitem{Kiryluk}
J.~Kiryluk, AIP Conf. Proc. {\bf 675}, 424 (2003).
\bibitem{Crawford}
H.~J.~Crawford, F.~S.~Bieser, T.~Stezelberger, {\it in preparation}.
\bibitem{Trueman_Kopeliovich} 
L.~Trueman, hep-ph/0203013 (2002); B.Z. Kopeliovich, AIP
Conf. Proc. {\bf 675}, 58 (2003).
\bibitem{Tojo}
J.~Tojo {\it et al.}, Phys. Rev. Lett. {\bf 89}, 052302 (2002).
\bibitem{Bravar}
A.~Bravar, {\it these Proceedings}.
\bibitem{CNI}
O.~Jinnouchi {\it et al.}, AIP Conf. Proc. {\bf 675}, 424 (2003).
\bibitem{vanderMeer}
A.~Drees and X.~Zu, {\it Proceedings of the Particle Accelerator
Conference 2001}, p. 3120.
\bibitem{STARpp}
J.~Adams {\it et al.}, Phys. Rev. Lett. {\bf 91}, 172302 (2003).
\bibitem{Brahms}
M.~Adamczyk {\it et al.}, Nucl. Instr. Meth. A {\bf 499}, 437 (2003).
\end{thebibliography}
\end{document}